\documentclass[sigconf,natbib=false]{acmart}
\AtBeginDocument{%
  }

\setcopyright{acmlicensed}
\copyrightyear{2026}
\acmYear{2026}
\acmDOI{XXXXXXX.XXXXXXX}

\acmConference[MLCAD '26]{International Symposium on Machine Learning for CAD}{September 7, 2026}{Jeju, South Korea}

\usepackage{booktabs}
\usepackage[utf8]{inputenc}
\usepackage{tabularx}
\usepackage{multirow}
\usepackage{float}
\usepackage{setspace}
\usepackage[
backend=biber,
style=numeric,
sorting=none
]{biblatex}
\copyrightyear{2026}
\acmYear{2026}
\setcopyright{cc}
\setcctype{by}
\acmConference[MLCAD '26]{2026 ACM/IEEE International Symposium on Machine Learning for CAD}{September 07--09, 2026}{Jeju Island, Republic of Korea}
\acmBooktitle{2026 ACM/IEEE International Symposium on Machine Learning for CAD (MLCAD '26), September 07--09, 2026, Jeju Island, Republic of Korea}
\acmDOI{10.1145/3831599.3840342}
\acmISBN{979-8-4007-2878-5/2026/09}
\begin{document}

\title{NoTB: Oracle-Free Triage of LLM-Generated RTL \\via Cross-Variant Formal Equivalence}
\title{NoTB: Oracle-Free Triage of LLM-Generated RTL \\via Cross-Model Formal Consensus}
\author{Elisavet Lydia Alvanaki}
\email{ealvanaki@cs.columbia.edu}
\affiliation{%
  \institution{Dept. of Computer Science\\Columbia University}
  \city{New York}
  \country{USA}
}
\author{Je Yang}
\email{je.yang@cs.columbia.edu}
\affiliation{%
  \institution{Dept. of Computer Science\\Columbia University}
  \city{New York}
  \country{USA}
}
\author{Biruk Seyoum}
\email{biruk@cs.columbia.edu}
\affiliation{%
  \institution{Dept. of Computer Science\\Columbia University}
  \city{New York}
  \country{USA}
}
\author{Luca P. Carloni}
\email{luca@cs.columbia.edu}
\affiliation{%
  \institution{Dept. of Computer Science\\Columbia University}
  \city{New York}
  \country{USA}
}








\begin{abstract}

Large language models (LLMs) are increasingly used to generate
register-transfer-level (RTL) designs from natural-language specifications.
However, assessing functional correctness at early stages remains a
fundamental challenge. Existing oracle-free approaches rely either on
simulation-based agreement, which depends on LLM-generated testbenches that
can fail or vary across models, or on LLM-as-a-judge heuristics, which
produce inconsistent predictions.

We introduce NoTB, an oracle-free triage framework that infers correctness
from cross-model formal consensus. NoTB generates RTL implementations from
multiple independently trained LLM families and applies Sequential
Equivalence Checking (SEC) to identify designs that are provably equivalent.
We show that the diversity of model families within an SEC-equivalent cluster
induces a calibrated correctness signal, enabling risk–coverage tradeoffs
without requiring testbenches.

On 78 CVDP RTL-generation tasks, four-family formal consensus achieves 94.7\% precision at 27\% coverage; three-family consensus achieves 87\% precision at 33\% coverage. These operating points give designers a tunable accept/defer rule before a trusted testbench or golden RTL is available. Overall, NoTB demonstrates that formal cross-model agreement provides a reliable basis for high-confidence triage without model-dependent oracles.

\end{abstract}





\maketitle

\section{Introduction}
\label{sec:intro}

Large language models (LLMs) are increasingly capable of generating synthesizable register-transfer-level (RTL) designs directly from natural-language specifications. Given a text description of desired hardware behavior, a model produces hardware-description language (HDL) code that is then compiled, simulated, and checked for correctness~\cite{liu2023verilogeval,thakur2023verilog,lu2024rtllm}.
Most existing methods rely on an \textit{oracle}, a trusted artifact, such as a reference testbench or golden RTL model, to evaluate the correctness early in the design process. Recent benchmarks report high pass rates against such oracles, suggesting that LLM-assisted hardware design is maturing~\cite{pinckney2025revisiting,yu2025spec2rtlagent}. 
In practice, however, constructing a reliable oracle is itself expensive and error-prone, and developers rarely have one at the start of a design task. 
A key challenge in LLM-assisted RTL design is therefore not candidate generation (models produce many candidates cheaply), but triage: deciding which candidates merit downstream verification before a trusted oracle becomes available.

Existing oracle-free approaches attempt to fill this gap, but remain fundamentally limited~\cite{zhao2025vrank,fang2025assertllm,zheng2023judging}. LLM-as-a-judge~\cite{zheng2023judging} replaces execution with model-based prediction, but provides no formal guarantees and exhibits strong model-dependent bias: false-acceptance rates vary by up to 45\% across generating models despite similar ground-truth pass rates, making it unreliable as a triage signal (as shown in Section~\ref{sec:why_formal}).
Operating on the intuition that the agreement of many models is evidence of correctness~\cite{wang2023selfconsistency,chen2023codet}, simulation-based self-consistency clusters implementations by behavioral agreement under a generated testbench.
This intuition rests on two strong assumptions that rarely hold in practice: that the testbench covers the relevant input space and that incorrect implementations disagree with each other. Both can fail simultaneously. A weak testbench can miss a critical input, thus allowing a false consensus to form among incorrect designs; this might collapse precision to 43\%, as shown in Section~\ref{sec:why_formal}. Moreover, simulation-based agreement compounds shared-error risk: when multiple
models converge on the same wrong behavior, a weak testbench can still falsely detect
that behavior as consensus~\cite{zhao2025vrank,chen2023codet}. These shortcomings reflect a fundamental limitation of existing heuristic approaches that infer correctness from incomplete or learned signals rather than from formal guarantees.

To address these challenges, we introduce \textbf{NoTB}, a framework that replaces a heuristic agreement with a \textit{formally certified agreement}. The central insight is that if implementations produced by independently trained LLM families\footnote{An LLM family is a set of LLMs derived from a common base model (e.g. through fine-tuning, distillation, or versioning) rather than trained independently from scratch. In this paper, we represent each LLM family by a single representative model.} are proven equivalent over the full input space via Sequential Equivalence Checking (SEC)~\cite{mneimneh2005sequential,pixley1992theory}, their agreement is a property of the designs, not of the testbench. Cross-family agreement is more likely to reflect correctness rather than shared error, since independently trained models exhibit different inductive biases and failure modes~\cite{wang2025codeerrors}, making convergence on the same wrong answer unlikely. 
SEC guarantees that agreement is not an artifact of a finite set of tests. 
Given a specification, NoTB generates implementations from multiple LLM families and applies SEC~\cite{mneimneh2005sequential,pixley1992theory} to identify sets of designs that provably are functionally identical throughout the input space. 
These equivalence classes partition the candidate set into clusters of interchangeable implementations. We define the \emph{model count} of a cluster as the number of distinct LLM families represented within it, and show that model count induces a calibrated correctness signal: a higher model count corresponds systematically to lower error rates. This enables risk-aware triage with explicit control over the precision--coverage\footnote{In this paper, \textit{coverage} denotes the percentage of specifications for which NoTB makes a prediction among those evaluated (e.g., out of all specifications in a given benchmark suite).}tradeoff. 
Crucially, NoTB does not replace verification. It selectively certifies the high-confidence subset so that those designs can bypass early-stage checks, while the remainder proceeds through the existing flow unchanged.

We evaluate NoTB on the CVDP benchmark~\cite{pinckney2025cvdp} using four independently trained LLM families. When the dominant equivalence cluster contains all four families, NoTB achieves 94.7\% precision at 27\% coverage; relaxing the threshold to three families yields 87\% precision at 33\% coverage. This monotonic relationship — stronger cross-model agreement, lower false-acceptance risk, is the central empirical result and enables users to select a confidence threshold suited to their verification budget. 

Our paper makes the following contributions:
\raggedbottom

\begin{itemize}
  \item We show that simulation-based oracle-free RTL triage is 
    testbench dependent: on a fixed candidate pool, the precision of  
    a four-way agreement varies from 43\% to 84\% purely as a 
    function of which LLM generated the testbench.

  \item We introduce NoTB, a novel framework that replaces the learned and 
    simulation-based agreement with formal cross-model consensus. 
    NoTB clusters candidate implementations by SEC and scores each cluster by its 
    \emph{model count}, i.e., the number of distinct LLM families 
    represented in the cluster.

  \item Our experiments with 78 CVDP RTL-generation tasks and  
    four LLM families demonstrate that model count is a calibrated 
    correctness signal: four-family agreement reaches 94.7\% 
    precision at 27\% coverage, and three-family agreement reaches 
    87\% at 33\%, with the signal remaining stable 
    under leave-one-out family ablation.
\end{itemize}
\raggedbottom

\section{Related Work}
\label{sec:related_work}
Oracle-free correctness assessment for LLM-generated RTL relies on proxy
signals in place of a trusted artifact. We group prior work by such signals:
learned judgment (LLM-as-a-judge), sample agreement (self-consistency), and
behavioral agreement under a generated testbench (simulation-based clustering).


\textbf{Heuristic Correctness Prediction (LLM-as-a-Judge).}
To reduce reliance on test execution, recent approaches adopt \emph{LLM-as-a-judge}~\cite{zheng2023judging}, where a secondary model predicts correctness given a specification and its RTL implementation. While this eliminates the need for testbenches, it replaces execution with a learned heuristic. Such predictions are inherently \emph{model-dependent} and can vary significantly between generating models, limiting their reliability as correctness signals. This variability introduces systematic bias, making LLM-based judgment unsuitable as a high-confidence triage signal.

\textbf{Agreement and Self-Consistency.}
Self-consistency methods use agreement among multiple samples as a
confidence signal~\cite{wang2023selfconsistency,li2022alphacode}. This
requires a meaningful notion of when two answers are the same. For RTL,
textual agreement is insufficient: syntactically different designs may
be equivalent, while similar designs may differ on corner-case
sequences. Hence, oracle-free RTL triage requires agreement on
hardware behavior, not on emitted code strings.

\textbf{Agreement under Incomplete Evaluation.}
Recent work has explored agreement-based selection for RTL generation~\cite{wang2023selfconsistency,li2022alphacode,}. In particular, VRank~\cite{zhao2025vrank} generates multiple candidate implementations and clusters them based on behavioral equivalence under a shared testbench, ranking clusters by consistency. However, agreement is defined with respect to a \emph{finite set of test inputs}. As a result, functionally distinct implementations may appear equivalent if differences are not exercised by the testbench. Moreover, clustering depends on the quality of the testbench itself, which is generated by an LLM, introducing an additional source of bias and uncertainty.



\textbf{Formal Agreement as a Correctness Signal.} 
NoTB differs from prior oracle-free methods in where agreement is
defined. LLM-as-a-judge defines agreement through a learned
evaluator. Simulation-based clustering defines agreement through a
generated testbench and a finite set of observed traces. NoTB defines
agreement through SEC: two implementations are clustered only
when a formal tool proves convergence to a common behavior.
\textit{
This distinction changes the role of agreement. In prior methods,
agreement is an empirical proxy for correctness. In NoTB, agreement
identifies a formally proven behavioral class, and model diversity
within that class is used as the empirical confidence signal.
}
\raggedbottom



\section{Proposed Methodology}
\label{sec:methodology}

\begin{figure}[t]
    \centering
    \includegraphics[width=0.9\linewidth]{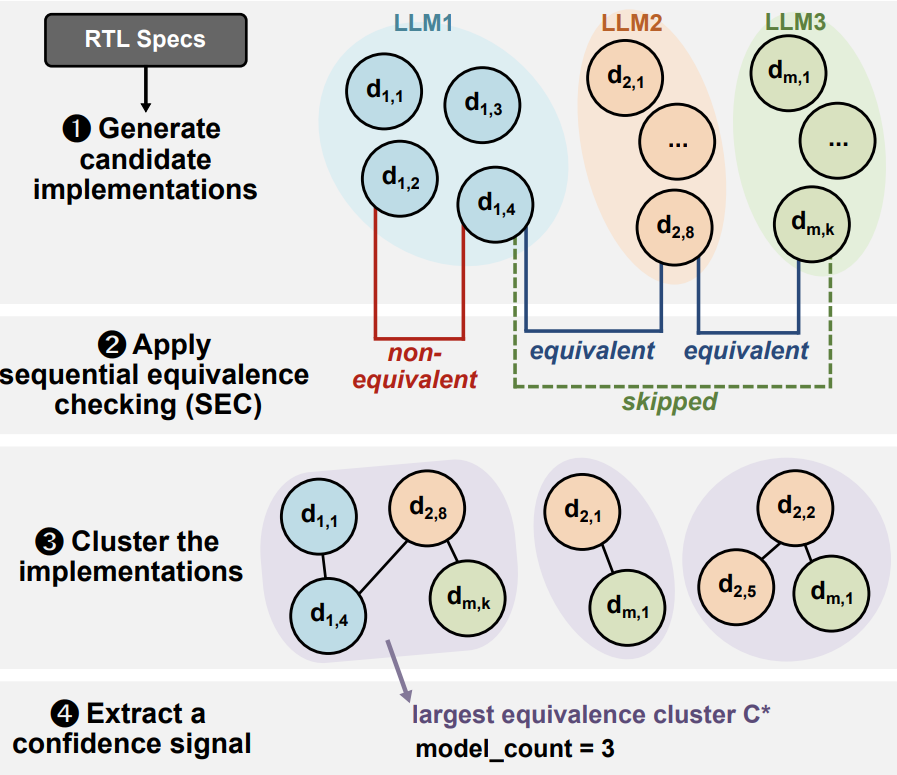}
    \vspace{-0.1in}
    \caption{Overall flow of the NoTB framework.}
    \vspace{-0.1in}
    \label{fig:overview}
\end{figure}


NoTB uses \emph{formal agreement among independent generators} as the triage
signal. If implementations produced by different LLM families are proven
equivalent for all input sequences, their agreement provides stronger evidence
than finite-test agreement or learned judgment. NoTB operationalizes this idea
in four stages, as shown in Figure ~\ref{fig:overview}. First, it samples RTL implementations from multiple LLM
families. Second, it applies SEC to compare
candidate implementations. Third, it constructs equivalence clusters from
proven SEC results. Finally, it scores the dominant cluster by its
\emph{model count}, defined as the number of distinct LLM families represented
in the cluster. This score enables a tunable precision--coverage tradeoff: increasing the threshold requires agreement among more model families, which reduces the number of accepted specifications but increases confidence that those accepted are correct.

\subsection{Multi-Model RTL Generation}

Given a specification $s$, NoTB constructs a candidate set by sampling from
multiple independently trained LLM families. Let
$\mathcal{M} = \{m_1, \dots, m_M\}$ denote the set of model families. For each
model $m \in \mathcal{M}$, we sample $K$ implementations, yielding
\[
\mathcal{D}(s) =
\bigcup_{m \in \mathcal{M}} \{d_{m,1}, \dots, d_{m,K}\}.
\]
Each implementation $d \in \mathcal{D}(s)$ is associated with its generating
model family $\mathrm{model}(d) \in \mathcal{M}$.

Sampling across model families, rather than only within a single model, exposes
different inductive biases and error modes. NoTB does not assume that any
individual model is reliable; it uses cross-family convergence as evidence.

All models are prompted to generate synthesizable Verilog under a shared set of
constraints: a single top-level module, no SystemVerilog-only constructs, no
simulation-only features such as \texttt{\small initial} blocks or delays, and a
single-driver discipline. These constraints standardize the generated
implementations and make them compatible with downstream SEC.

\subsection{Formal Equivalence Clustering}

Given the candidate set $\mathcal{D}(s)$, NoTB constructs a graph that captures
formal equivalence relationships among implementations. For each pair
$(d_i,d_j)$, we invoke Cadence JasperGold to determine whether the two designs
produce identical outputs for all input sequences. To automate comparisons
across independently generated RTL, NoTB infers clock and reset signals from
port names such as \texttt{\small clk}, \texttt{\small reset}, and \texttt{\small rst\_n}.

Each SEC query yields one of three outcomes: \emph{equivalent},
\emph{non-equivalent}, or \emph{inconclusive}. We define $d_i \equiv d_j$ only
when equivalence is proven. Tool errors, interface mismatches, and inconclusive
results do not contribute equivalence edges.

NoTB represents the SEC results as an undirected graph

$G(s) = (V,E)$, where $V = \mathcal{D}(s)$ and
$(d_i,d_j) \in E \iff d_i \equiv d_j$.

The connected components of $G(s)$ define the equivalence clusters $\mathcal{C}(s) = \{C_1,\dots,C_k\}.$
Each cluster corresponds to a set of implementations connected through proven
equivalence. Equivalence may be direct or transitive: if $d_i \equiv d_j$ and
$d_j \equiv d_k$ are both proven, then $d_i$, $d_j$, and $d_k$ belong to the
same cluster. However, a direct graph edge between $d_i$ and $d_k$ is added only
if their pairwise SEC query also proves equivalence. Thus, clusters use
transitive closure, but edges remain faithful to individual SEC proofs.

This construction is conservative. NoTB merges implementations only when
equivalence has been formally established. Non-equivalence results help separate
behaviors, and inconclusive results leave candidates unmerged. Hence, each
cluster represents a behavior shared by implementations that are provably
equivalent under the SEC model.

\subsection{Incremental SEC Pruning}
\label{subsec:eff_discovery}
Naively evaluating all pairs requires $O(|\mathcal{D}(s)|^2)$ SEC queries, or
$O(M^2K^2)$ queries for $M$ model families and $K$ samples per family. NoTB
reduces this cost using lightweight pruning while preserving the invariant that
equivalence edges are introduced only by proof.

First, NoTB eliminates syntactically identical implementations through hashing.
Identical files are merged into the same cluster before SEC. Second, NoTB
maintains equivalence components using union-find. When SEC proves
$d_i \equiv d_j$, their components are merged, and future comparisons within
the resulting cluster are skipped. NoTB also records proven non-equivalences
between component representatives to avoid redundant cross-component
comparisons.

These optimizations affect which comparisons are skipped, not which
equivalences are trusted. A cluster is still formed only from implementations
connected by proven SEC equivalence.

\subsection{Confidence as Model Diversity}

Given the equivalence clusters $\mathcal{C}(s)$ of a specification $s$, NoTB derives a correctness
signal from model diversity. For each cluster $C \in \mathcal{C}(s)$, we define
its \emph{model count} as the number of distinct model families represented in
the cluster:
\[
\mathrm{model\_count}(C) =
\left|\{\mathrm{model}(d) : d \in C\}\right|.
\]

Let
\[
C^* = \arg\max_{C \in \mathcal{C}(s)} |C|
\]
denote the largest equivalence cluster. 
NoTB assigns specification $s$ the confidence score
\[
\mathrm{confidence}(s) = \mathrm{model\_count}(C^*).
\]
The largest cluster is treated as the dominant behavioral hypothesis for the
specification. Its model count measures how many independent model families
converged to that behavior.

If multiple clusters tie in size, NoTB selects the cluster with higher model
count. Remaining ties are broken deterministically using a fixed ordering over
cluster members . These tie-breaking rules affect only ambiguous cases and do not
change the definition of the confidence signal.

Given a threshold $\tau$, NoTB accepts specification $s$ if
\[
\mathrm{confidence}(s) \ge \tau.
\]
Varying $\tau$ induces a precision--coverage tradeoff: higher thresholds accept
fewer specifications but require stronger cross-model formal consensus. 
This formulates correctness estimation as a selective prediction.
Hence, NoTB is a selective-prediction method, not a universal
correctness classifier. It accepts only the specifications whose
dominant formally equivalent behavior has sufficient cross-family
support, and defers the rest to the existing verification flow.
\section{Evaluation}
\label{sec:experimental}
We evaluated NoTB on precision and coverage metrics with an experimental setup that consists of datasets, models, and oracle-free baselines.

\begin{figure}[t]
    \centering
    \includegraphics[width=\linewidth]{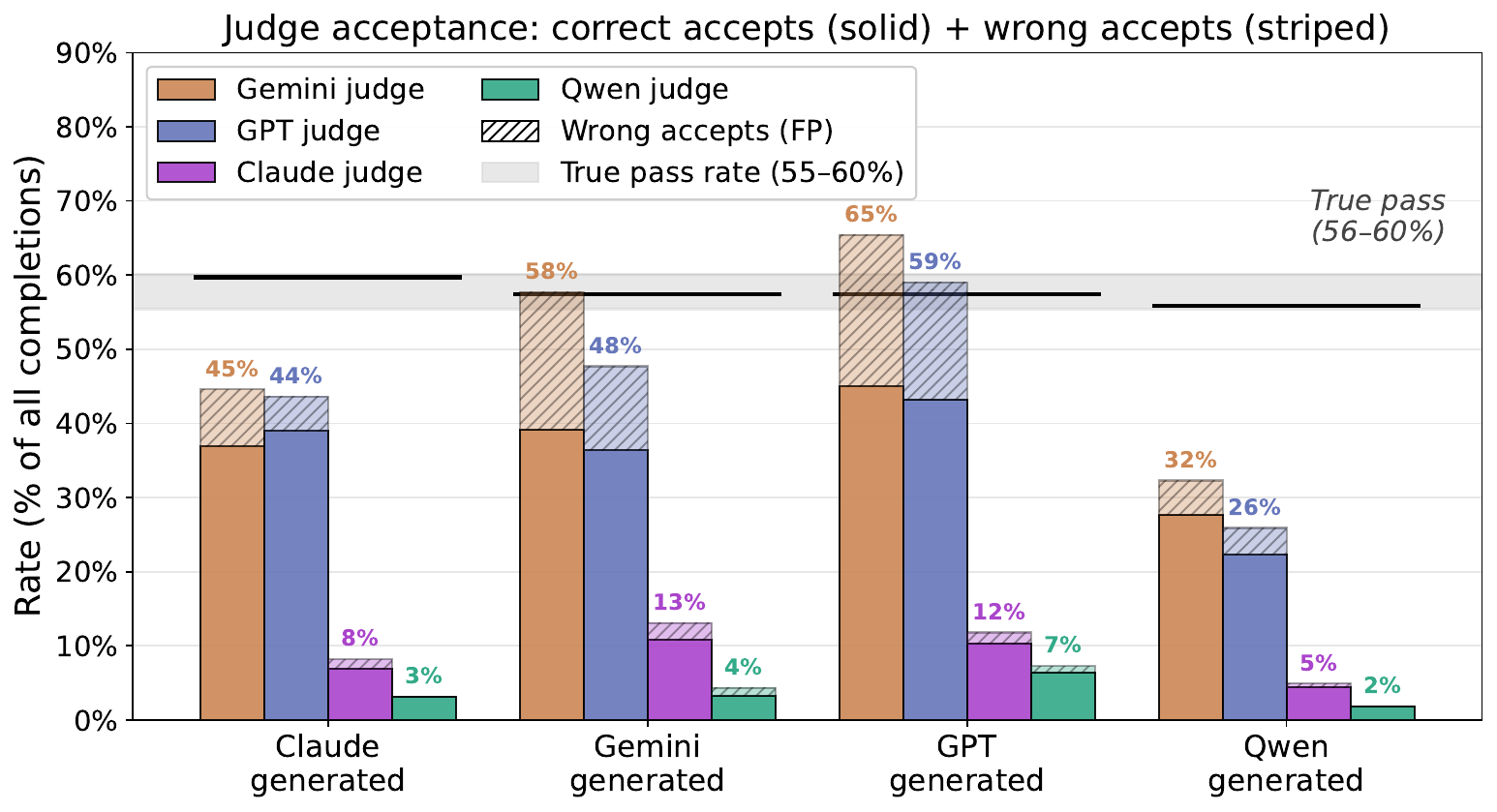}    
    \vspace{-17pt}
    \caption{Judge acceptance decomposed into correct accepts (solid) and false
    accepts (striped), for four judges across four generating models. The dashed
    line marks the ground-truth pass rate.}
    \label{fig:judge_model_dependence}
    \vspace{-10pt}
\end{figure}
\begin{figure*}[!t]
    \centering
    \includegraphics[width=0.8\textwidth]{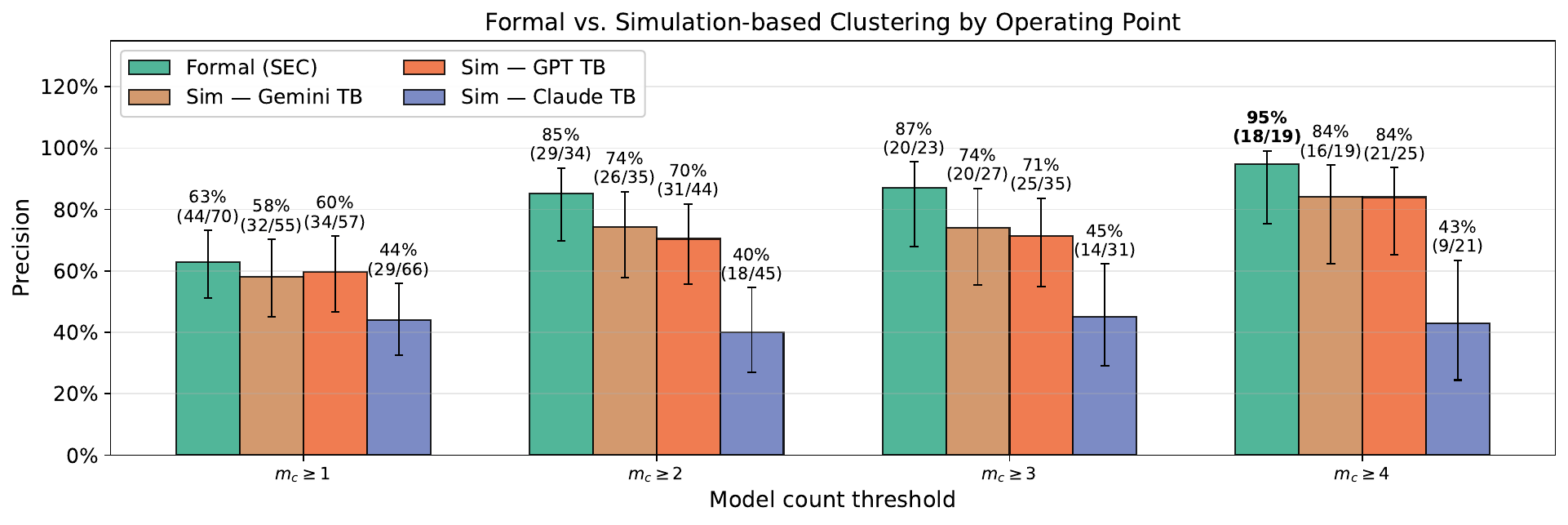}
    \vspace{-13pt}
    \caption{Precision at each model-count threshold for formal SEC and three
    simulation-based clusterings. Counts show correct/accepted specifications
    per bar.}
    \label{fig:coverage_vs_sim_cumulative}
    \vspace{-8pt}
\end{figure*}

\subsection{Experimental Setup}

\paragraph{Dataset and models.}
We evaluate NoTB on all 78 code-generation hardware design tasks from the
code-generation tasks (\texttt{cid003}) of the NVIDIA CVDP
benchmark~\cite{pinckney2025cvdp}. For each specification, we generate
candidate RTL implementations using $M=4$ independently trained LLM families:
Claude 3.7 Sonnet~\cite{Claude3S}, GPT-OSS-120B~\cite{openai2025gptoss120bgptoss20bmodel}, Gemini 2.5 Flash~\cite{comanici2025gemini25pushingfrontier}, and Qwen 3 Coder~\cite{Qwen3-Coder-Next}. We sample
$K=5$ implementations per model, yielding up to 20 candidates per specification.


 \vspace{-2pt}
We use the official \textit{cocotb} testbenches solely to establish ground truth, not to construct NoTB's triage signal. A selected cluster is counted as correct only if every implementation it contains passes all functional tests. In NoTB, members of the same SEC cluster are functionally identical under the cocotb testbenches, consistent with their proven equivalence. Therefore, if one implementation fails under a given test, other implementations in the equivalence cluster do the same.

We report functional correctness after excluding compile-time and spec-level failures, all of which are independent of NoTB's signal  as they are determined by ordinary compilation and applied uniformly across all methods. Of the 78 tasks, we remove 5 on which Yosys fails to elaborate the design and the test harness cannot run, 2 for which fewer than 4 candidate implementations are synthesizable, leaving too small a pool for cross-model consensus, and 1 whose LLM-generated testbench fails to compile or deadlocks. In summary, we have 70 specifications evaluated across all models. We construct equivalence clusters using pairwise SEC with JasperGold, following the methodology in Section ~\ref{sec:methodology}.


\paragraph{Baselines.}
We compare against two oracle-free baselines. First,
\emph{LLM-as-a-judge}~\cite{zheng2023judging} uses an LLM to predict
correctness from the specification and implementation. We evaluate four judges
and aggregate predictions at the specification level. Second,
\emph{simulation-based consensus} clusters implementations by agreement under
generated testbenches~\cite{zhao2025vrank}.

\paragraph{Metrics.}
We report precision (correct accepts / total accepts) and coverage 
(accepts / evaluable specs) at the specification level, with 95\% Wilson 
confidence intervals~\cite{wilson1927probable}. When reported, false positive (FP) rate 
normalizes false positives by the total number of incorrect specifications, 
recall by the total number of correct ones, and accuracy is the fraction of 
predictions matching ground truth. Statistical significance is assessed 
using Fisher's exact test when comparing high- and low-consensus groups.

\subsection{Why Formal Consensus?}
\label{sec:why_formal}
We quantify the instability of LLM-as-a-judge- and testbench-dependent oracle-free baselines:
 \vspace{-4pt}
\subsubsection{LLM-as-a-Judge Ablation}
We evaluate LLM-as-a-judge as a correctness proxy using four judge models:
Gemini 2.5 Flash~\cite{comanici2025gemini25pushingfrontier}, GPT-OSS-120B~\cite{openai2025gptoss120bgptoss20bmodel}, Claude 3.7 Sonnet~\cite{Claude3S}, and Qwen 3 Coder~\cite{Qwen3-Coder-Next}.

\begin{table}[t]
\centering
\setlength{\tabcolsep}{4pt}
\small
\caption{Cross-judge ablation on CVDP. Results measured at completion level. Bold denotes best per column.}
    \vspace{-8pt}
\label{tab:judge_cross_model}
\begin{tabular}{ll rrrr}
\toprule
\textbf{Judge} & \textbf{Gen.\ model} & \textbf{FPR\,(\%)} & \textbf{Rec\,(\%)} & \textbf{Prec\,(\%)} & \textbf{Acc\,(\%)} \\
\midrule
Gemini 2.5 Flash & Claude & 19.1 & 61.8 & 82.8 & 69.5 \\
 & Gemini & 43.4 & 68.3 & 68.0 & 63.3 \\
 & GPT & 47.9 & 78.3 & 68.8 & 67.2 \\
 & Qwen & 10.5 & 49.5 & 85.7 & 67.2 \\
 & \emph{Overall} & 29.4 & \textbf{64.0} & 74.7 & 66.8 \\
\midrule
Claude 3.7 Sonnet & Claude & 3.2 & 11.6 & 84.4 & 45.9 \\
 & Gemini & 5.4 & 18.8 & 82.4 & 51.0 \\
 & GPT & 3.6 & 18.0 & 87.2 & 51.4 \\
 & Qwen & 1.2 & 7.8 & 89.5 & 47.9 \\
 & \emph{Overall} & 3.3 & 13.9 & 85.1 & 49.0 \\
\midrule
GPT-OSS-120B & Claude & 11.5 & 65.2 & 89.4 & 74.6 \\
 & Gemini & 26.5 & 63.4 & 76.3 & 67.7 \\
 & GPT & 37.1 & 75.1 & 73.2 & 69.9 \\
 & Qwen & 8.1 & 39.9 & 86.1 & 62.8 \\
 & \emph{Overall} & 20.2 & 60.5 & 80.3 & \textbf{68.7} \\
\midrule
Qwen 3 Coder & Claude & 0.0 & 5.2 & 100.0 & 43.3 \\
 & Gemini & 2.4 & 5.8 & 76.5 & 44.9 \\
 & GPT & 2.1 & 11.1 & 87.5 & 48.0 \\
 & Qwen & 0.0 & 3.2 & 100.0 & 45.9 \\
 & \emph{Overall} & \textbf{1.1} & 6.1 & \textbf{88.3} & 45.4 \\
\bottomrule
    \vspace{-17pt}
\end{tabular}
\end{table}

Although ground-truth pass rates are similar across generating models
(55.9--59.7\%), judge acceptance varies substantially with the code
generator: Gemini and GPT each exhibit a 33-point spread across
generators (Figure~\ref{fig:judge_model_dependence}). The cross-judge
ablation in Table~\ref{tab:judge_cross_model} shows that this instability
persists across judges: higher-recall judges incur substantial
false-positive rates, while conservative judges reduce false positives
primarily by abstaining. Thus, LLM-as-a-judge does not provide a stable
high-confidence triage signal.

\subsubsection{Simulation-Based Clustering is Testbench-Dependent}


To isolate the effect of the equivalence mechanism, we hold the RTL 
candidate pool fixed and vary only the generated testbench used for 
clustering. On the same candidate pool, simulation-based precision 
at model count (mc)$\geq$4 varies from 43\% with Claude-generated testbenches to 
84\% with Gemini- and GPT-generated testbenches, while formal SEC 
reaches 94.7\% (Figure~\ref{fig:coverage_vs_sim_cumulative}). 
Simulation agreement is therefore partly a property of the generated 
testbench, not only of the RTL candidates.

The following failure modes emerged as part of our analysis: For the 64b/66b encoder task 
(specification ID~5), Claude-generated tests place four CVDP-failing 
candidates and sixteen passing candidates in the same mc=4 cluster 
because the stimulus does not exercise the distinguishing behavior. 
On specification ID~122, a GPT-generated testbench deadlocks after reset 
and produces no clusters despite all candidates compiling. SEC merges 
candidates only under proven equivalence and requires no stimulus, 
avoiding both failure modes.

Together, these results motivate formal consensus as the confidence 
signal. Judge-based and simulation-based agreement are properties of 
the evaluator and testbench, respectively, rather than of the RTL 
candidates. NoTB instead defines agreement through SEC, so cross-model 
consensus reflects the designs themselves.

\subsection{NoTB Performance}
\label{sec:formal_consensus}

\subsubsection{Triage Performance}

We evaluate whether cross-model agreement under formal equivalence predicts
functional correctness. For each specification, we measure the model diversity
of the largest equivalence cluster and analyze how this signal relates to
correctness. Figure~\ref{fig:coverage_vs_sim_cumulative} and
Table~\ref{tab:model-count}  summarize the result.


\begin{table}[t]
\caption{Selective prediction as a function of model count. Each row defines a decision
rule: accept specifications whose largest cluster contains at least the
indicated number of model families.}
\vspace{-5pt}
\label{tab:model-count}
\centering
\small
\begin{tabular}{lcrrrcc}
\toprule
Threshold & Acc. & Corr. & FP & Precision & 95\% CI & Coverage \\
\midrule
$\geq 4$ & 19 & 18 & \textbf{1} & \textbf{95\%} & [0.75, 0.99] & 27\% \\
$\geq 3$ & 23 & 20 & 3 & 87\% & [0.68, 0.95] & 33\% \\
$\geq 2$ & 34 & 29 & 5 & 85\% & [0.70, 0.94] & 49\% \\
All      & 70 & 44 & 26 & 63\% & [0.51, 0.73] & 100\% \\
\bottomrule
\end{tabular}
\vspace{0.3em}

\end{table}

As shown in Table~\ref{tab:model-count}, correctness improves with model count ($mc$). 
At the lowest threshold, mc$\geq1$, precision is 63\%, reflecting the baseline difficulty of the benchmark.
Increasing the threshold to mc$\geq2$ and mc$\geq3$ raises precision to 85.3\% and 87\%, respectively. 
At full four-family agreement, mc$\geq4$, precision reaches 94.7\%. This trend shows that model
diversity within the dominant SEC cluster provides a calibrated correctness
signal: stronger formal cross-model agreement corresponds to lower empirical
error risk. 

SEC turns agreement into a property of the RTL implementations, rather than of a finite testbench. NoTB therefore prioritizes high-consensus clusters while leaving lower-consensus cases in the standard validation flow. The mc$\geq4$ regime is the highest-precision operating point, accepting
19 specifications at 94.7\% precision and 27\% coverage. The mc$\geq3$ regime is
a broader operating point: it increases coverage to 33\% while maintaining 87\%
precision. This behavior is well aligned with early-stage RTL triage, where the
goal is not to classify every generated implementation, but to identify the
subset of specifications whose dominant behavior has enough formal cross-model
support to advance with lower false-acceptance risk.
The separation between high-consensus (mc$\geq4$) and low-consensus (mc$<4$) regimes is statistically
significant. Their comparison yields
$p=4.4\times10^{-4}$ under Fisher's exact test. Taken together, these results
show that formal cross-model agreement provides a reliable and calibrated basis
for early-stage RTL triage.

\subsubsection{Model Family Sensitivity}

To assess whether any single model family is load-bearing for the consensus
signal, we re-cluster each specification using only three of the four model
families, dropping all five completions from the omitted family.
Table~\ref{tab:loo} reports precision and coverage at each model-count threshold
for each leave-one-out setting.

\begin{table}[t]
\centering
\vspace{-5pt}
\caption{Leave-one-out ablation. Each row drops one model family (5 completions) and re-clusters the remaining 3~families. mc$\geq$3 means all models agree.}
\label{tab:loo}
\begin{tabular}{lrrrrrr}
\toprule
Dropped & \multicolumn{2}{c}{mc$\geq$3} & \multicolumn{2}{c}{mc$\geq$2} & \multicolumn{2}{c}{mc$\geq$1} \\
model   & Prec & Cov & Prec & Cov & Prec & Cov \\
\midrule
  Claude & 90\% & 29\% & 82\% & 40\% & 63\% & 100\% \\
  Gemini & 95\% & 27\% & 83\% & 43\% & 63\% & 100\% \\
  GPT & 95\% & 27\% & 89\% & 40\% & 63\% & 100\% \\
  Qwen & 100\% & 29\% & 81\% & 46\% & 63\% & 100\% \\
\bottomrule
\end{tabular}
\vspace{-10pt}
\end{table}

Dropping Qwen leaves mc$\geq3$ precision at 100\%. Dropping Claude reduces mc$\geq3$ precision to 90\%, indicating that Claude completions provide useful discriminative
coverage. Importantly, high-precision triage remains possible in every
leave-one-out setting. NoTB's confidence signal therefore does not depend on
access to any particular model family, making the method portable across
deployment settings with different model availability.
\subsubsection{Completions per Model}

\begin{figure}[t]
    \centering
    \includegraphics[width=\linewidth]{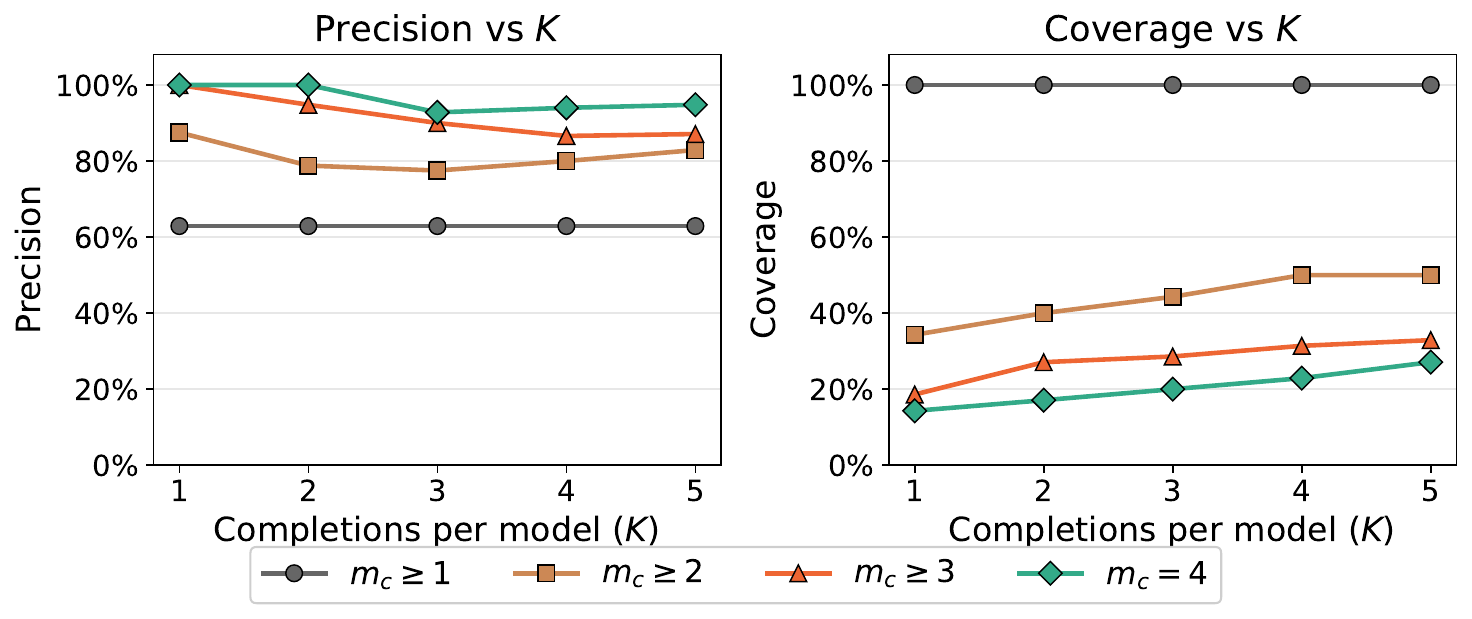}
    \caption{Precision and coverage at each model-count threshold as $K$ varies
    from 1 to 5.}
    \label{fig:k_curve}
    \vspace{-10pt}
\end{figure}

We vary the number of completions (generated RTL samples) per model family, $K \in \{1,2,3,4,5\}$,
and recompute model count using the corresponding formal-equivalence clusters
(Figure~\ref{fig:k_curve}).
At mc$\geq4$, precision remains above 93\% across the sweep, while coverage
increases from 14\% at $K=1$ to 27\% at $K=5$. Gains diminish beyond $K=3$,
suggesting that a modest number of samples per family captures most
high-confidence cases.
\vspace{-10pt}
\subsubsection{Cost Analysis}

\begin{table}[h]
\caption{NoTB pipeline cost vs.\ baseline costs. All costs computed from measured
per-call token rates at current OpenRouter list pricing.}
\vspace{-5pt}
\label{tab:cost}
\centering
\small
\begin{tabular}{llrr}
\toprule
Stage & Component & Calls & Cost \\
\midrule
\multirow{4}{*}{Generation}
            & Claude 3.7 Sonnet & 390 & \$12.29  \\
            & Gemini 2.5 Flash                         & 390 & \$1.88   \\
            & Qwen 3 Coder                             & 390 & \$0.59   \\
            & GPT-OSS-120B                      & 380 & \$0.17   \\
\midrule
\multirow{3}{*}{TB Gen.}
            & Claude 3.7 Sonnet & 77 TBs & \$10.52 \\
            & Gemini 2.5 Flash  & 78 TBs & \$3.64  \\
            & GPT-OSS-120B & 78 TBs & \$0.00 \\
\midrule
\multirow{4}{*}{Judge}
            & Claude 3.7 Sonnet & 1{,}556 & \$27.99 \\
            & Gemini 2.5 Flash  & 1{,}556 & \$27.89 \\
            & Qwen 3 Coder      & 1{,}556 & \$5.00  \\
            & GPT-OSS-120B & 1{,}556 & \$0.00 \\
\midrule
SEC         & JasperGold        & 13{,}133 pairs & license \\
\bottomrule
\end{tabular}
 \vspace{-5pt}

\footnotesize

\end{table}

We quantify the monetary and computational cost of NoTB across all 78
specifications. Table~\ref{tab:cost} summarizes candidate generation,
judge-based evaluation, and formal equivalence.

\begin{figure}[h]
    \centering
    \includegraphics[width=\linewidth]{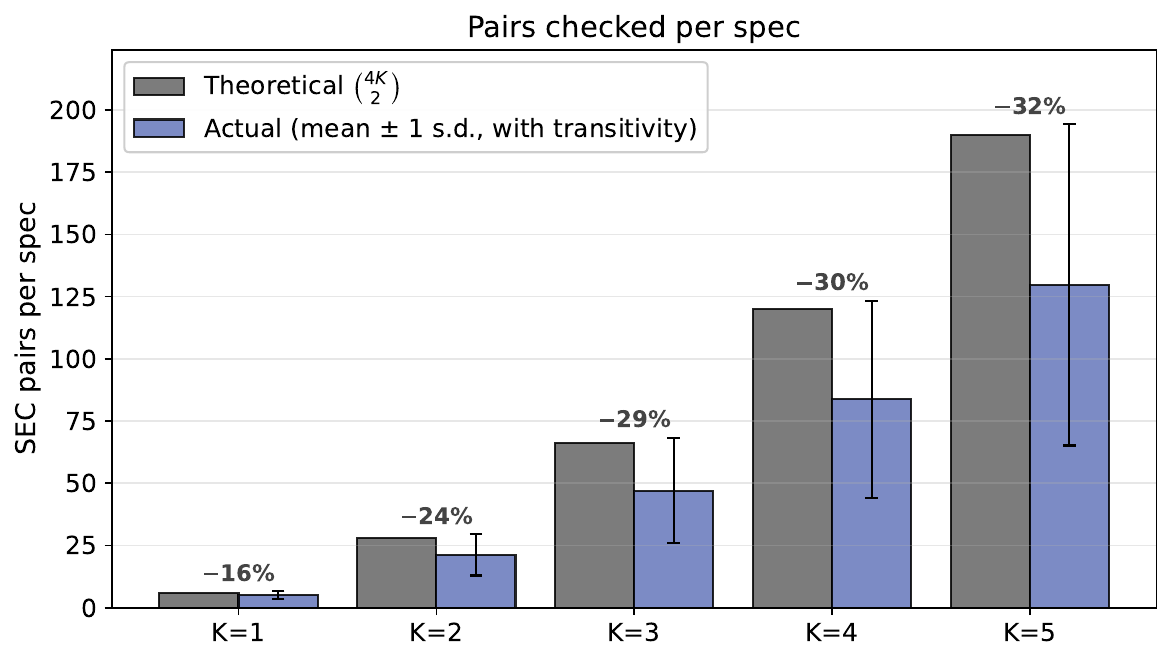}
    \vspace{-20pt}
    \caption{SEC pairs per specification: theoretical worst case
    $\binom{4K}{2}$ versus actual count after hash deduplication and
    transitivity pruning. Error bars show standard deviation across
    specifications.}
    \label{fig:sec_pairs}
    \vspace{-7pt}
\end{figure}

\begin{figure}[h]
    \centering
    \includegraphics[width=0.95\linewidth]{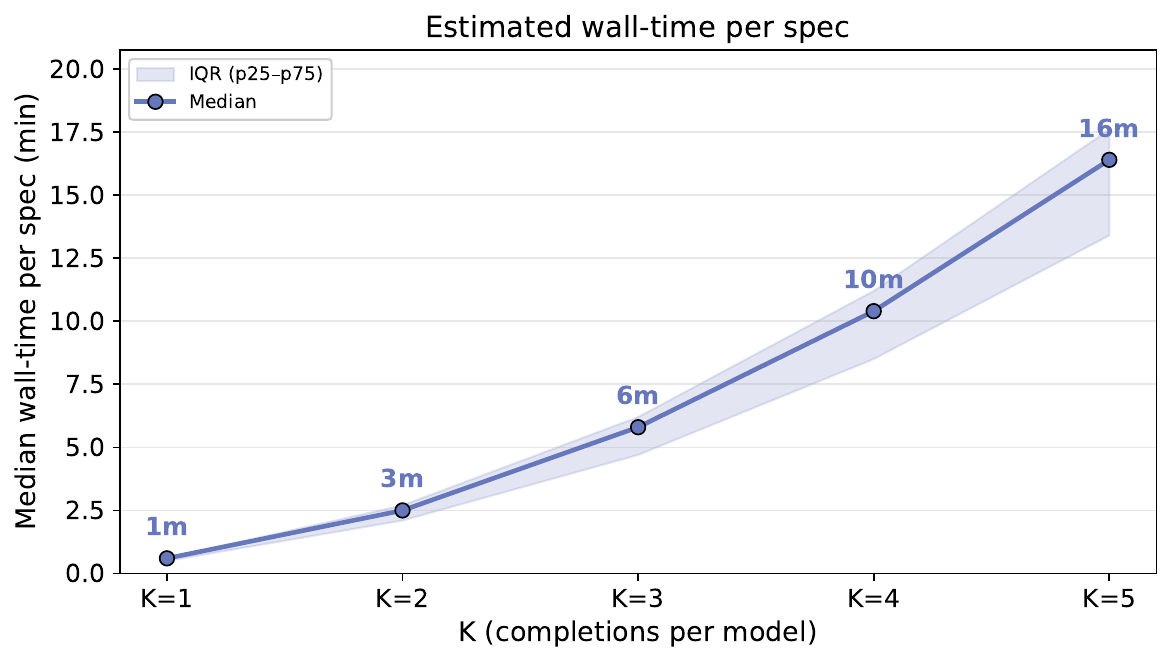}
    \vspace{-10pt}
    \caption{Median wall-clock time per specification as $K$ varies. The shaded
    region shows the 25th--75th percentile.}
    \label{fig:sec_walltime}
    \vspace{-10pt}
\end{figure}

Candidate generation is shared by all evaluated methods. Generating 20
implementations per specification across four model families costs \$14.93 over
the full dataset. NoTB then performs triage using local SEC checks. It does not
require additional judge calls or generated testbenches to construct its
confidence signal. In contrast, the LLM-as-a-judge baseline requires 1,556
additional model calls in our evaluation.

Without pruning, comparing all candidate pairs would require up to $\binom{20}{2}=190$ SEC invocations per specification. Hash-based deduplication and transitivity pruning reduce this by roughly 32\% on average, though savings vary across specifications (Figure~\ref{fig:sec_pairs}). Across the dataset, the majority of SEC calls yield definitive equivalent or non-equivalent verdicts, with the remainder terminating during elaboration or returning inconclusive. Only proven equivalences contribute to clusters, so inconclusive results reduce coverage but cannot introduce false merges.

Median wall-time per specification grows from 1 minute at $K=1$ to 16 minutes at
$K=5$ (Figure~\ref{fig:sec_walltime}). SEC calls are independent across pairs
and specifications, so the pipeline is parallelizable. Overall, NoTB trades the per-call LLM spend of judge- and testbench-based baselines for local SEC compute that parallelizes across pairs; the baselines' costs scale linearly with candidates and specifications, while NoTB's reduce to wall-clock on existing verification hardware.
\vspace{-15pt}
\section{Conclusion}

NoTB is an oracle-free triage framework
for LLM-generated RTL that replaces testbench- and judge-based 
agreement with formal cross-model consensus. Our approach leverages Sequential Equivalence Checking to cluster 
candidate implementations using behavioral agreement over the full input space, and scores each cluster by the number 
of distinct LLM families represented in it. NoTB 
outperforms existing oracle-free baselines by achieving up to 
94.7\% precision at 27\% coverage on 78 CVDP RTL-generation 
tasks across four LLM families.

\begin{acks}
This work was supported in part by Amazon.com, Inc.
\end{acks}
\printbibliography

\end{document}